\documentclass[12 pt]{article}
\def\baselinestretch{1.0}
\catcode`@=11
\@addtoreset{equation}{section}
\begin{document}
\begin{center}
{ \def\baselinestretch{2.0}
\Large \bf  MULTIPLE ATTRACTOR IN NEWTON-LEIPNIK SYSTEM, PEAK-TO-PEAK DYNAMICS AND CHAOS CONTROL }
\end{center}

\vskip 20pt
\begin{center}
{\it {   Biswambhar Rakshit\footnote {e-mail: biswa\_ju2004@hotmail.com}, Papri Saha\footnote {e-mail: papri\_saha@yahoo.com} and A. Roy Chowdhury$^*$\footnote{$^*$(corresponding author) e-mail: arcphy@cal2.vsnl.net.in }  }\\
                           High Energy Physics Division \\
                             Department Of Physics     \\
                             Jadavpur University        \\
                              Kolkata - 700032\\
                                    India \\}
\end{center}
\vskip 30pt
\begin{center}
\underline{\large Abstract}
\end{center}
	The chaotic properties of Newton-Leipnik system are discussed from the view point of strange attractors.  Previously, two strange attractors of this system were illustrated which occured from two different initial conditions under the same parameter condition. It is found that above system also exhibits multiple attractors under different parameter values but same  initial condition and we have shown the existence of three other  strange attractors with  varying dimensionality under different parametric conditions. The properties of these attractors are then analyzed on the basis of Lyapunov exponents, power spectra, recurrence analysis and peak-to-peak dynamics. The peak-to-peak dynamics  relies on the low dimensionality of the chaotic attractor and allows to approximately model the system. Peak-to-peak plot along with return-time plot are then effectively used to solve the optimal control problem of the system which reverts the system to a periodic situation.
\newpage
\section{Introduction}
Beginning with the pioneering work of Lorenz on convectional instabilities [1], there has been a lot of research work on systems of differential equations which yield bounded nonperiodic trajectories converging to an attractor of rather complicated nature. A chaotic attractor has a kind of ``hank of string" like appearance and does not have a integral dimension. The chaotic trajectory is not one-dimensional as it does not return over itself, nor is it two-dimensional because it is not a surface uniformly covered by trajectories, nor is it a volume uniformly filled by the trajectories to become a three-dimensional structure.\par
In this communication, we have studied the Newton-Leipnik chaotic system [2] which possesses more than one attractor. Newton and  Leipnik obtained the  set of  differential equations from Euler rigid body equations which were modified with the addition of a linear feedback. Two strange attractors starting from different  initial conditions and same parameter conditions were obtained. B. Marlin established the existence of closed orbits which were not asymptotically stable [3]. Chen et al.[4]  devised a stable-manifold-based method for chaos control and synchronization. H. Richter studied the stabilization of a desired motion  within one attractor as well as taking the system dynamics from one attractor to another[5].
\par
We have shown that  when the parametric conditions are varied keeping the initial conditions intact, the system showed the presence of multiple attractors. The chaoticity of these attractors are then analyzed through phase space analysis, power spectrum[6,7] and lyapunov exponents[8,9]. Dimension of the attractor which is fractal is estimated. Owing to the  low dimensionality of the attractors, the peak-to-peak dynamics[10,11] is analyzed and a suitable controller [12,13]has been designed for the system which reduces it to a purely periodic state. 
\section{Formulation}
\subsection{Analysis of Chaos}Newton Leipnik system is characterised by the following differential equations.
\begin{equation}
\dot{x_1}=-\alpha x_1+x_2+10 x_2 x_3
\end{equation}
\begin{equation}
\dot{x_2}=- x_1-0.4 x_2+5 x_1 x_3
\end{equation}
\begin{equation}
\dot{x_3}=\beta x_3-5 x_1 x_2
\end{equation}
This system has two strange attractors originating from the initial states $(0.349, 0.0,  -0.160)^T$, $(0.349,  0.0,  -0.180) ^T$ and system parameters $\alpha =0.4, \beta =0.175,$ which  have been studied and analyzed previously[2,4].\par But while exploring the dynamical system with respect to the parameters we came across three other chaotic attractors. These three chaotic attractors emerged from a totally different parameter state and initial condition. The two dimensional projection of these attractors are shown in figure 1(a), 1(c), 1(e) and their corresponding   power spectrum is illustrated in figure 1(b), 1(d), 1(f)  respectively. The initial values of the variable, in our case, is taken as $ x_{10}=0.35$,  $x_{20}=0.0$, $x_{30}=0.16$. The parameter $`\beta '$ is kept constant at 0.17 whereas $`\alpha '$ is varied. The attractor in figure 1(a),  appears for $\alpha =0.550$ whereas that of  figure 1(c) is for $\alpha =0.7689$ and the last attractor of figure 1(e) appears at $\alpha =0.809$.  Changing of the parameter $`\alpha '$ causes a  rapid change in the appearance of the strange attractor and also  in its nonlinear properties.\par The first attractor is quite but not exactly similar to the one described in [4]. The second one represents much like a chaotic torus. The third one is carrying an entirely  different structure. It is also clear from the power spectrum also that in comparison to the other two the second  figure shows a little bit ordered behaviour which is also observable in the phase space diagram. We have also studied the lyapunov spectrum, which forms the most useful tool for the detection and quantification of chaos. Lyapunov exponents are the average exponential rates of divergence or convergence of nearby orbits in phase space. Any system containing atleast one positive lyapunov exponent is defined to be chaotic, with the magnitude of the exponent reflecting the  time scale on which  system dyanamics becomes unpredictable. The lyapunov exponents calculated for the attractors are tabulated as follows.\\
\begin{center}
{\bf TABLE I}\\
\vskip 10pt
\begin{tabular}{|c|c|c|c|} \hline
{}& {}& {}& {}\\
 & $\lambda _1$ &$\lambda _2$&$\lambda _3$\\
{}& {}& {}& {}\\
\hline
{}& {}& {}& {}\\
Attractor 1&0.25554&0.03268&-1.18911\\
{}& {}& {}& {}\\
\hline
{}& {}& {}& {}\\
Attractor 2&0.18986&0.1039005&-1.19466\\
{}& {}& {}& {}\\
\hline
{}& {}& {}& {}\\
Attractor 3&0.21343&0.15888&-1.27319\\
{}& {}& {}& {}\\
\hline
\end{tabular}
\end{center}
The Lyapunov spectrum of a strange attractor is closely associated to its fractional dimension. Kaplan and Yorke [14--17]  conjenctured that the information dimension $d_f $ is related to the lyapunov spectrum by the equation
$$ d_f = j + \frac{\sum_{i=1}^j \lambda _i}{\mid \lambda _{j+1}\mid }, $$
where $j$ is defined by the condition that
$$\sum_{i=1}^j\lambda_i >0   \; \;\mbox{and}\;\;\sum_{i=1}^{j+1}\lambda_1<0   $$
The information dimension for the three attractors are obtained as,
                        $$d_{f_1}=2.24 \;\;\;\;\;\; d_{f_2}=2.25 \;\;\; \;\;\;d_{f_3}=2.29$$
It is seen that with the increase in parameter  $`\alpha '$, the information dimension also increases. From the lyapunov spectrum it is observed that the largest positive lyapunov decreases in the case of the second attractor and again rises for the third. So the expansion rate of the trajectories of the second attractor is less than that of the other two attractors whereas the third attractor displays the  largest  information dimension.
  
\par After three attractors have been displayed with different value of parameter $`\alpha '$ we explore the whole spectrum of the system with  the variation of  $`\alpha'$.  This is shown by means of  the bifurcation diagram of figure (2).  There is period doubling route  to chaos which is also depicted through phase space diagrams shown in figure (3a)-(3d). The first figure shows a single period with $\alpha =0.8045$. As we decrease the parameter value of $\alpha $ to 0.798 the phase space shows a double period. Further decrement of $\alpha $ to 0.7944 quadrapules the period which switches to a chaotic situation after a few more doublings and at a value $\alpha =0.7689$.
\par \par After this analysis on the chaotic dynamics of the system, we have taken the analysis of the peak-to-peak dynamics of the system which has been further used to design a controller for controlling the system. 
\subsection{Peak-to-peak Dynamics}
It was first shown by E.N. Lorenz that for a chaotic attractor the intensity of the forthcoming peak and its time of occurrence can be predicted with sufficient accuracy by one or more curves in the plane $(x_i, x_{i+1})$, where $x_i$ denotes a maxima in the temporal evolution of the variable $X(t)$. This characteristic of the attractor depends on the lower dimensionality of the chaotic attractor, which must be near about two. The subject of peak- to-peak dynamics have been discussed in the various field of science and engineering. R\"{o}ssler system, the Chua circuit and the Duffing oscillator are found to display peak-to-peak dynamics under suitable parameter conditions. A system which display peak-to-peak dynamics can be described by a reduced order model which is the peak-to-peak map and a return time map. The former predicts the value of the next peak, while the second one gives the time interval separating a peak from the next peak. The methodology of obtaining a peak-to-peak plot is that, from the time series of a particular variable $X$, one collects the peak points as $x_i$ and plots them as $x_i$ vs $x_{i+1}$. Thus the plot is composed of distinct points. The attractor is periodic if the PPP shows only a few distinct  points. The points of the PPP are all distinct and may show a definite curve if it is a case of torus or strange attractor with a dimension close to two. On the other hand for a high dimensional strange attractor one obtains a cloud like set. For a quasi-periodic orbit the points of PPP form a closed regular curve. In the similar way a return time plot can be obtained from the set of all pairs $(x_i, \tau_i)$ with $\tau_i= t_{i+1}-t_i$. \par
The peak-to-peak plot and the return time plot for the three attractors are depicted in figures 4(a)--4(f). The PPP and RTP for the first attractor, which occurs for $\alpha =0.55$ and has a fractal dimension $2.24$, shows a well behaved structure as shown in figure 4(a) and 4(b). 4(c) and 4(d) is for the attractor with $\alpha =0.7689$ and fractal dimension $2.25$ and a distinct curve is obtained. 4(e) and 4(f) is for the attractor with $\alpha =0.809$ and fractal dimension $2.29$ and this is reflected in the PPP and RTP which has a cloud like appearance. This indicates that peak-to-peak dynamics is dependent on the low dimensionality of the system and is clearly visible from the above analysis.
\subsection{Control of Chaos}
The idea of control of chaos was enunciated by Ott, Grebogi and Yorke [22] and after that various methodologies and strategies have been developed theoretically and experimental realization of control have been achieved [23]. Control of chaos have been effectively observed in a magnetoelastic ribbon [24], a heart [25], a thermal convection
loop [26], a yttrium iron garnet oscillator [27], a diode oscillator [28], an optical multimode
chaotic solid-state laser [29], a Belousov-Zhabotinski reaction diffusion chemical system [30], and
many other experiments.
\par Peak-to-Peak plot and return time plot can be used to device optimal feedback controllers of chaotic systems. If the PPP is a smooth curve then one can attain a map from it. In our observation the attractor with $\alpha =0.7689$ showed a perfect curve (parabola) and it has been targeted to achieve the controlled state.  This control procedure is adopted easily once the peak-to-peak map is derived. Our analysis has been considered in an ideal environment where noise is absent and the sampling has been made on sufficiently high frequency. In our system we introduce a control variable $u(t)$ in the third equation as follows:
\begin{equation}
\dot{x_1}=-\alpha x_1+x_2+10 x_2 x_3
\end{equation}
\begin{equation}
\dot{x_2}=- x_1-0.4 x_2+5 x_1 x_3
\end{equation}
\begin{equation}
\dot{x_3}=\beta x_3-5 x_1 x_2-bu(t)
\end{equation}
where b is a real constant to be introduced later. For any constant control $u(t) = \bar u \in U = [u_{\mbox{min}}, u_{\mbox{max}}]  \subset R$  the system has a chaotic attractor. In our case for $u(t)=0.0$,  three chaotic attractor were obtained. In order to control the system, one of the requisites is that the control law $u(t)$ is constant between two subsequent peaks i.e., $u(t)=u_k \;\;\forall \;\;t \in (t_k, t_{k+1}]$. We consider the third state variable as output variable.
A peak-to-peak map $$x_{k+1}=F(x_k, \bar u)$$ have been derived, through simulation, for 20 equally spaced values of $u$ in the set $U = [0.0, 0.0095]$. 
The reduced order model
\begin{equation}
x_{k+1}=F(x_k,u_k)
\end{equation}
  have been obtained by linear interpolation of these maps. Thus, the dynamics of variable $y$ of a system can be easily described by equation (2.7). The fixed point        $$\bar x=F(\bar x,\bar u),$$  identifies a  unstable periodic orbit of 
the system $$ \dot{x}(t)=f(x(t),u(t)) $$
The output $x(t)$ associated with the unstable periodic orbit has one peak per period at $\bar x$. The linear approximation of the map $F$ around $\bar x$ is given by
$$x_{k+1}-\bar x=A(x_k-\bar x) + b(u_k-\bar u)$$
where,
$$A=\left[\frac{\partial F}{\partial x}\right]_{\bar x, \bar u}\;\;\;\;\;\mbox{and}\;\;\;\;\; b=\left[ \frac{\partial F}{\partial u}\right]_{\bar x, \bar u} $$
 are two real numbers that we assume to be not equal to zero. The law for control is then given by
$$u_k= \left\{ \begin{array}{c} \bar u+ h(x_k- \bar x) \;\;\;\;\; \mbox{if} \mid x_k- \bar x\mid \leq \epsilon \\
\\
\bar u \;\;\;\;\;\;\;\;\hskip 50pt \mbox{if} \mid x_k- \bar x\mid > \epsilon  \end{array}\right.$$
where the scalar $h$ must be such that $ \mid A+bh <1\mid $. So the control $u_k$ is allowed to vary from $(\bar u-h \epsilon )$ to $(\bar u+h \epsilon )$ and
$$ h=-\frac{A}{b}\;\;\;\; \Rightarrow \;\;\;\;A+bh = 0$$
is optimal because the response time of the linear approximation is minimum (i.e., equal to one), since, $x_{k+1}= \bar x$.\par
In the case of Newton Leipnik System, at $\bar u=0.007$  the  PPM  $x_{k+1}=F(x_k, \bar u)$ has an unstable fixed point at $\bar x=0.3$ with $A=-3.6$ , $b=0.601$. The nature of the controlled system with $h=-A/b=6.0$ ,  $\epsilon=0.5$ is shown in figure (5). The nature of the phase space is depicted in the inset  that  represents a single periodic orbit.
\section{Conclusion}
Three new attractors of the  Newton Leipnik system have been analyzed, which was found to evolve upon changing the parameter condition of the system, keeping the initial conditions fixed. Observation shows that they not only differ in structure and properties but also show varying dimensionality. Peak-to-peak dynamics have been obtained  and the peak-to-peak plot is found to be useful in determining the controller function which reverts the chaotic situation to a periodic state.\\
\\
{\bf Acknowledgement:}\\
Authors, P. Saha and B. Rakshit, are thankful to CSIR, Govt. of India for a senior research fellowship and a junior research fellowship respectively. They would also like to thank Dr. Benjamin Marlin of North Western Oklahoma State University, for providing the background of the literature and many helpful references.
\section{References:}
{[1]}.  E.N. Lorenz - {\it J. Atmos. Sci.} {\bf \underline {20}}  {(1963)} {130}.\\
{[2]}.  R.B. Leipnik and T.A. Newton - {\it Phys. Lett. A} {\bf \underline {86}}  {(1981)} {63}.\\
{[3]}.  B.A. Marlin - {\it Int. J. Bifurc. \& Chaos} {\bf \underline {12}}  {(2002)} {511}.\\
{[4]}.  S. Chen et. al. - {\it Chaos, Solitons \& Fractals} {\bf \underline {20}}  {(2004)} {947}.\\
{[2]}.  H. Richter - {\it Phys. Lett. A} {\bf \underline {300}}  {(2002)} {182}.\\
{[6]}. D.Farmer et.al - {\it Ann.N.Y.Acad.Sci.}{\bf\underline{357}} {(1980)} {453}\\
{[7]}. R.J.Higgins - {\it Ann.J.Phys.}{\bf\underline {44}} {(1976)} {766}\\
{[8]}. A.Wolf , J.Swift , H.Swineny and J.Vastano - {\it Physica D} {\bf \underline {16}}  {(1985)}{285}\\
{[9]}. Hao.Bai-lin , Elementary Symbolic Dyanmics and Chaos in Dissipative Systems.(World Scintific , Singapore , 1989).\\
{[10]}.  M. Candaten and S. Rinaldi - {\it Int. J. Bifurc. \& Chaos} {\bf \underline {8}}  {(2000)} {1805}.\\
{[11]}.  C. Piccardi and S. Rinaldi - {\it Int. J. Bifurc. \& Chaos} {\bf \underline {6}}  {(2003)} {1579}.\\
{[12]}.  C. Piccardi and S. Rinaldi - {\it Int. J. Bifurc. \& Chaos} {\bf \underline {12}}  {(2002)} {2927}.\\
{[13]}.  C. Piccardi and S. Rinaldi - {\it Physica D} {\bf \underline {144}}  {(2000)} {298}.\\
{[14]}. J. Kaplan and J. Yorke, ``Chaotic behaviour of multidimensional difference equations"  in Functional Differential Equations and the approximation of fixed points, lecture Notes in Mathematics, vol 730.  H.O.peitgen and H.O. Walther eds. (Springer, Berlin) p.228. \\                   
{[15]}. P.Frederickson, J.kaplan, E.Yorke and J.Yorke 
%"The lyapunov dimension of strange attractors" %J.Diff Eqs 49(1983)185.
- {\it J. Diff. Eqs. } {\bf\underline{49}} {(1983)} {185}\\
{[16]}. L. S. Young - {\it Ergodic theory and Dyanamical Systems } {\bf\underline{2}} {(1982)} {109}\\
{[17]}. F.Ledrappier - {\it Comm. Math. Phys.} {\bf\underline{81}} {(1981)} {229}\\
{[18]}.  J.P. Eckmann, S.O. Kamphorst and D. Ruelle  - {\it Europhys. Lett.} {\bf \underline {4}}  {(1987)} {973}.\\
{[19]}.  A. Giuliani and C. Manetti - {\it Phys. Rev. E} {\bf \underline {53}}  {(1996)} {6336}.\\
{[20]}.  J.P. Zbilut and C.L. Webber. Jr. - {\it Phys. Lett. A} {\bf \underline {171}}  {(1992)} {199}.\\
{[21]}. P. Grassberger and I. Procaccia - {\it Physica D } {\bf \underline{7}} {(1983)} {473}\\
{[22]}. E. Ott, C. Grebogi, J.A. Yorke - {\it Phys. Rev. Lett. } {\bf \underline{ 64 }} {(1990} {1196}\\
{[23]}. S. Boccaletti et al. - {\it Physics Reports} {\bf \underline{329}} {(2000)} {103}\\
{[24]}. W.L. Ditto,  S.N. Rauseo and M.L. Spano - {\it Phys. Rev. Lett. } {\bf \underline{65}} {(1990)} {3211}\\
{[25]}. A. Garfinkel et al.- {\it Science } {\bf \underline{257}} {(1992)} {1230}\\
{[26]}. J. Singer and H.H. Bau - {\it Phys. Fluids A} {\bf \underline{3}} {(1991)} {2859}\\
{[27]}. A. Azevedo and S. Rezende - {\it Phys. Rev. Lett.} {\bf \underline{66}} {(1991)} {1342}\\
{[28]}. E. R. Hunt - {\it Phys. Rev. Lett. } {\bf \underline{67}} {(1991)} {1953}\\
{[29]}. R. Roy et al. - {\it Phys. Rev. Lett. } {\bf \underline{68}} {(1992)} {1259}\\
{[30]}. V. Petrov et al. - {\it Nature } {\bf \underline{361}} {(1993)} {240}\\

\section{Figure Captions}
Figure (1) - 2D phase space diagram and the coresponding power spectrum for Newton-Leipnik System. (a) \& (d) : $\alpha =0.55$   (b) \& (e) :  $\alpha =0.7689$ (c) \& (f) :  $\alpha =0.809$.\\
\\
Figure (2) - Bifurcation diagram of the system with respect to the variation of the parameter $\alpha $.\\
\\
Figure (3) - Period doubling route to chaos. (a) : $\alpha =0.8045$, (b) : $\alpha =0.798$, (c) : $\alpha =0.7944$, (d) : $\alpha =0.7689$.\\
\\
Figure (4) - Peak -to-peak plot and return time plot. (a) \& (b): $\alpha =0.55$,  (c) \& (d): $\alpha =0.7689$, (e) \& (f): $\alpha =0.809$.\\
\\
Figure (5) - Variation of $x_3(t)$ with the application of the controller showing transition to a periodic state. Inset picture shows the phase space diagram during the controlled stage.

\end{document}